\documentclass[8pt]{article}

\usepackage[table]{xcolor}
\usepackage{graphicx,pslatex,float,color,array,amssymb,amsmath,euscript}
\usepackage{pxfonts}
\usepackage{ifthen}
\usepackage{longtable}
\usepackage{eso-pic}
\usepackage{sidecap}
\usepackage{xspace}
\usepackage[paperwidth=21.0cm,paperheight=29.7cm,textwidth=15cm,textheight=20cm,top=2cm,left=3.5cm,bottom=2cm]{geometry}
\graphicspath{{}}

\definecolor{navy}{rgb}{0,0,0.5}
\usepackage[pdftex,
           hyperindex=true,
           colorlinks=true,
           linkcolor=navy,
           anchorcolor=magenta,
           citecolor=navy,
           urlcolor=navy,
           unicode,
           implicit=true]{hyperref}

\renewcommand{\vec}[1]{\boldsymbol{#1}}
\newcommand{\be}{\begin{equation}}
\newcommand{\ee}{\end{equation}}
\newcommand{\bes}{\begin{equation*}}
\newcommand{\ees}{\end{equation*}}
\newcommand{\atan}{\mathrm{atan}}

\title{On the accurate computation of the true contact-area in mechanical contact of random rough surfaces}

\author{Vladislav A. Yastrebov$^a$\footnote{Corresponding author $\langle$\texttt{vladislav.yastrebov@mines-paristech.fr}$\rangle$}, Guillaume Anciaux$^b$, Jean-Fran\c cois Molinari$^b$}

\date{\small{\it$^a$MINES ParisTech, PSL Research University, Centre des Mat\'eriaux}\\{\it CNRS UMR 7633, BP 87, 91003, Evry, France}\\
\small{\it
$^b$Civil Engineering Department, Materials Science Department,}\\{\it Ecole Polytechnique F{\'e}d{\'e}rale de Lausanne (EPFL), Station 18, 1015, Lausanne, Switzerland}}

\begin{document}

\maketitle

\begin{flushleft}
 \large{\bf Abstract.}\normalsize
\end{flushleft}

\noindent We introduce a corrective function to compensate errors in contact area computations coming from mesh discretization. 
The correction is based on geometrical arguments and requires only one additional quantity to be computed: the length of contact/non-contact interfaces. 
The new technique enables us to evaluate accurately the true contact area using a coarse mesh for which the shortest wavelength in the surface spectrum reaches the grid size. 
The validity of the approach is demonstrated for surfaces with different fractal dimensions and different spectral content using a properly designed mesh convergence test.
In addition, we use a topology preserving smoothing technique to adjust the morphology of contact clusters obtained with a coarse grid.\\[4pt]

\begin{flushleft}
 {\bf Keywords:} \normalsize roughness, contact, simulations, boundary element method 
\end{flushleft}

\section{Introduction}

The roughness of natural and industrial surfaces determines properties of the mechanical contact between solids: interface stiffness, true contact area and the morphology of the free interface volume. Thus the roughness governs many interface phenomena such as contact electrical resistance~\cite{Timsit2013Slade,yastrebov2015holm}, thermal contact resistance~\cite{mikic1966thermal,anciaux2013ijhmt}, friction~\cite{dieterich1994direct,ben2010dynamics,rice2006heating}, adhesion~\cite{fuller1975effect,pastewka2014sticky}, wear~\cite{bowden2001b,aghababaei2016critical}, as well as fluid transport at contact interfaces~\cite{muller1998fluid,persson2005jpcm,sahlin2008lubrication,dapp2012prl,dapp2016fluid,rafols2016modelling}. 
For most of these phenomena it is critical to accurately estimate the true contact area for given thermo-electro-mechanical loads and given roughness of contacting surfaces. It is now well known that a simple load bearing area model, relying on a geometrical overlap of two rough surfaces considerably overestimates the true contact area~\cite{pei2005jmps,ramisetti2011autocorrelation}, and for equal contact areas, the former results in a much higher transmissivity in transport problems~\cite{dapp2012prl}. Existing analytical models, asperity-based~\cite{greenwood1966prcl,bush1975w,thomas1999b,mccool1986w,greenwood2006w,afferrante2012w}, as well as Persson's model~\cite{persson2001jcp,persson2002prb} with its adjusted version~\cite{persson2006contact}, rely on a few approximations and thus cannot provide very accurate results in terms of true contact area over a wide interval of loading conditions (for a detailed discussion and comparison see~\cite{manners2006w,carbone2008jmps,paggi2010w,yastrebov2012pre,yastrebov2015ijss}). 

For these reasons, a numerical analysis, free of restrictive assumptions, is now widely used for the study of rough contact: the finite element method~\cite{pei2005jmps,yastrebov2011cras}, a wide class of continuum boundary element methods~\cite{stanley1997,polonsky1999numerical,polonsky2000fast,liu2002studying}, discrete methods based on molecular dynamics~\cite{campana2006prb,campana2007epl,yang2006epje} or basic molecular dynamics~\cite{akarapu2011prl,spijker2013ti,pastewka2014sticky}. 
Continuum models are particularly attractive since they permit to cover a large spectrum of length scales. However, they are subject to discretization and convergence errors. The former is related to the finite size of the used grid/mesh, whereas the latter is related to the strong non-linearity and discontinuous nature~\cite{KikuchiOden,DuvautLions2} of contact problems requiring iterative procedures (Newton-Raphson method, iterative solvers) to achieve convergence, or in the case of explicit techniques for both finite element~\cite{hyun2004pre} and the Green's function molecular dynamics~\cite{campana2006prb,campana2007epl} obtaining the results requires damping of elastic vibrations. Various continuous numerical methods in contact mechanics exist~\cite{Wriggers2006}, among them penalty and barrier methods, Lagrange multipliers, augmented Lagrangian and other techniques, which convert the constrained optimization problem to an unconstrained one (or at least partially unconstrained). Some of these methods allow accurate satisfaction of contact constraints for a given discretization whereas others (like penalty or barrier type method) only approach the exact solution with an  accuracy that depends on the choice of parameters. In addition, different contact discretization techniques, which integrate contact tractions in the weak form in the finite-element method,
provide varying accuracy and convergence rates. These depend on the interpolation order of elements, mesh densities on contacting surfaces and mesh curvature. For details, see~\cite{Wriggers2006,Puso2004,Fri04,Wohlmuth,popp2010dual,yastrebov2013numerical,drouet2015optimal} and references therein.

Assuming that the convergence is ensured and that the numerical method is accurate enough, the discretization remains the sole limiting point in achieving accurate results in contact problems.
From experience, it is known that for non-conformal but simple geometries (1D or 2D wave, circle or sphere against a flat surface), a rather dense mesh is required at the contact interface to track the contact area evolution, see e.g.~\cite{mesarovic1999spherical,mesarovic2000frictionless,jackson2005finite,du2007finite,yastrebov2011cras,song2013elastic}. We also refer to a study of a bi-wavy surface in contact with a rigid flat~\cite{yastrebov2014tl}, in which a very dense grid (4096 points per side per wavelength) was used, which allowed the authors to reveal peculiar mean contact pressure behaviour near the percolation point, which was missed in previous studies. Extrapolating to the case of ``rough'' surfaces, we need to ensure accurate discretization of all harmonics present in the surface spectrum, i.e. the shortest wavelength $\lambda_s$ should be sufficiently discretized to ensure accurate estimation of the contact area. This statement implies that the surface ``roughness'' should be smooth enough (or should be smoothly interpolated from experimental data ~\cite{hyun2007ti,yastrebov2011cras}) and the ratio $\lambda_s/\Delta x$, where $\Delta x$ is the grid step, should be kept rather big~\cite{yastrebov2012pre,prodanov2014tl}.
Importantly, as was demonstrated in~\cite{yastrebov2015ijss}, the discretization error is affected not only by the shortest wavelength $\lambda_s$ but also by the 
longest wavelength $\lambda_l$ in the spectrum\footnote{For surfaces with a plateau in the surface spectrum, $\lambda_l$ corresponds to the shortest wavelength present in the plateau as in examples shown in Fig.~\ref{fig:rough}(c-e).}, since it determines the number of individual macro-contact clusters (see~\cite{persson2004nature,yastrebov2015ijss}).  Thus, since the error in the true contact area is proportional to the length of the boundary between contact and non-contact zones, a shorter longer-wavelength, results in more contact clusters and leads to a higher discretization error.
Hence, for any rough surface with a sufficiently large discrete spectrum, the discretization requirement may rapidly become a bottle neck in terms of computational resources.
Two alternative solutions can be used: sacrifice the accuracy by using a coarse mesh and/or use new more efficient numerical methods~\cite{polonsky1999numerical,bemporad2015optimization} and adapted hardware combining CPU and GPU and/or parallel algorithms. A notable example, of combining a Green's function molecular dynamics with computations on GPU enables researchers to make rough contact simulations on a grid with more than 17 billion(!) grid points~\cite{prodanov2014tl}.

The true contact area in numerical simulations of contact can be computed as the total area of surface-faces being in contact plus the areas associated with nodes on the contact-non contact border. In spectral methods~\cite{stanley1997,liu2002studying} using FFT techniques and in discrete techniques~\cite{campana2006prb,campana2007epl} requiring regular discretizations (equally spaced grid points), the true contact area fraction can be simply computed as a ratio between all points in contact (points with zero gap and non-zero pressure) to the total number of points. However, this estimation of the contact area slightly overestimates the true solution. The coarser the discretization, the higher the error: the convergence rate with element size is linear~\cite{yastrebov2015ijss}.

Here, we suggest an alternative approach that allows an estimation of the true contact area with high accuracy on a reasonably coarse mesh, i.e. enabling full representation of the surface spectrum. The approach is based on a corrective function which uses the length of the contact/non-contact boundary (or simply the perimeter of contact clusters). 
The method was already introduced in~\cite{yastrebov2015ijss}, but the previous study lacked a mesh convergence analysis and the corrective factor was not evaluated. Here, we correct these shortcomings and demonstrate the accuracy of the suggested techniques on numerous cases.

The paper is organized as follows: in Section~\ref{sec:method}, all numerical methods and models are presented: 
a method used to generate model roughness in Subsection~\ref{ssec:roughness},
a spectral method used to solve contact mechanical problems in Subsection~\ref{ssec:fft}.
The area-correction method is described in detail in Subsections~\ref{ssec:contact_area} and \ref{ssec:factor}.
In Section~\ref{sec:mesh_conv}, we present mesh convergence tests and demonstrate the performance of the approach.
A morphology smoothing of contact clusters is briefly outlined in Subsection~\ref{sec:morphology}.
A short conclusion is drawn in Section~\ref{sec:conclusion}.

\section{Methods \label{sec:method}}

\subsection{Rough surface generation \label{ssec:roughness}}

\begin{figure}[htb!]
  \includegraphics[width=1\textwidth]{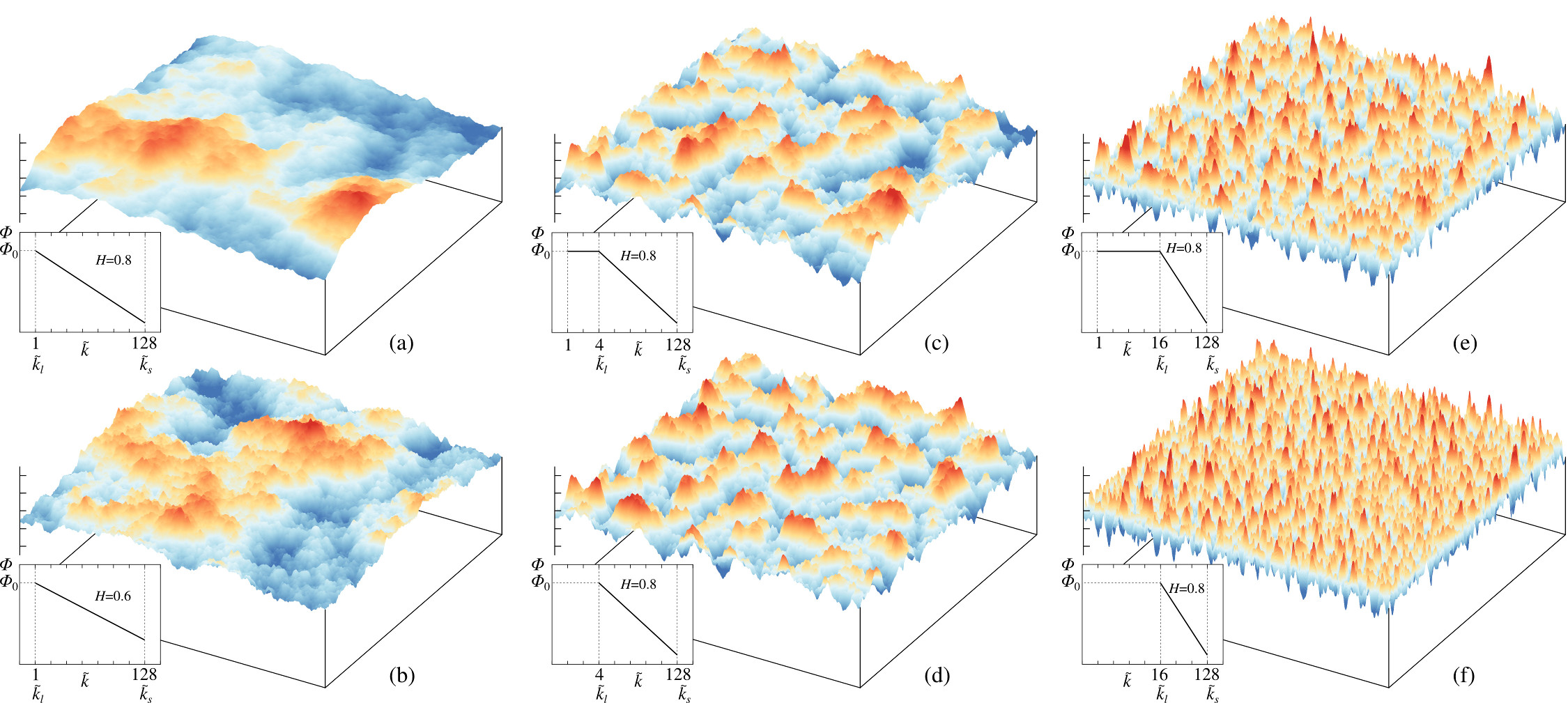}
  \caption{\label{fig:rough}Synthetic rough surfaces and corresponding power spectral densities: (a,b) $\tilde k_l=1$; (c) with and (d) without plateau for $\tilde k_l=4$;  (e) with and (f) without plateau for $\tilde k_l=8$; for all surfaces $\tilde k_s=128$. To show better the differences between surfaces with and without plateau, all surfaces for $H=0.8$ are generated by filtering from the same white noise.}
\end{figure}

We use an FFT filtering technique~\cite{hu1992ijmtm} to synthesize an isotropic 2D random rough surface with prescribed Hurst exponent $H$ (or, equivalently, fractal dimension $D=3-H$) and cut-offs in the surface spectrum as was done in~\cite{yastrebov2012pre,yastrebov2015ijss}. Note that we use a periodic surface resulting in a discrete spectrum. Surfaces with and without plateau in the power spectral density (PSD) are considered (see~\cite{jacobs2016arxiv} for a detailed discussion of PSD measurements and interpretation). One can compare surfaces with plateau depicted in Fig.~\ref{fig:rough}(c),(e) with those without it, see Fig.~\ref{fig:rough}(d),(f). 
The surface PSD is characterized by the following parameters: a scaling factor $\Phi_0$, Hurst exponent $H$, absence or presence of a plateau and also two wavenumbers $k_l$ and $k_s$:
$k_l$ the lowest wavenumber determining the start of a power-law decaying PSD and $k_s$ the highest wavenumber determining the shortest wavelength present in the spectrum.
The mean profile PSD\footnote{Lower index ``p'' is used for profile PSD and lower index ``s'' is used for surface spectrum.} for a surface with and without plateau can be formalized in the following forms, respectively:
\be
\Phi_p(k) = \begin{cases} \Phi_0, &\mbox{ if } k \le k_l\;\mbox{ (plateau) }\\ \Phi_0 (k/k_l)^{-2(H+1)}, &\mbox{ if } k_l < k \le k_s\\ 0, &\mbox{ else}\end{cases},
\qquad
\Phi_p(k) = \begin{cases} \Phi_0 (k/k_l)^{-2(H+1)}, &\mbox{ if } k_l < k \le k_s\\ 0, &\mbox{ else}\end{cases},
\label{eq:rough_descr}
\ee
where $k = \sqrt{k_x^2+k_y^2}$, with $k_x,k_y$ are wavenumbers in $x$ and $y$ directions ($k_{x,y} = 2\pi/\lambda_{x,y}$), respectively, and 
$$\Phi_p(k) = \frac{1}{\pi} \int\limits_0^{\pi} \Phi_s(k_x,k_y) \,d\phi = \frac{1}{\pi} \int\limits_0^{\pi} \Phi_s(k\cos(\phi),k\sin(\phi)) \,d\phi$$ 
is the mean profile PSD for a given absolute value of the wavevector $k$ in all directions determined by $\phi$. The surface is considered to be isotropic, so that statistical characteristics of every profile should be independent of the direction $\phi$.
Since we deal with a discrete spectrum the integral in the previous expression should be rewritten as:
$$
\Phi_p(k) = \frac{1}{n(k)}\sum\limits_{\forall k_x,k_y\;:\;k_x^2+k_y^2 = k^2} \Phi_s(k_x,k_y),
$$
where $n(k)$ is the number of elements satisfying $k_x^2+k_y^2 = k^2$ for all $k_x,k_y \in \mathbb Z$.
Because of this discreteness, the accuracy of definition~\eqref{eq:rough_descr} cannot be satisfied for a single surface. 
Thus, multiple realizations of rough surfaces for a given set of parameters $k_l,k_s,H$ is needed to capture the statistical nature of the roughness.
It is convenient to introduce dimensionless wavenumbers $\tilde k$, which will be used throughout the paper:
$$
  \tilde k = kL/2\pi = L/\lambda,
$$
where $\lambda$ is the wavelength and $L$ is the length of a side of simulation box. It means that $\tilde k$ is the number of waves per length, and because of the periodicity this quantity takes only integer values.
Some examples of generated rough surfaces with and without plateau are depicted in Fig.~\ref{fig:rough}.

\subsection{Spectral boundary element method\label{ssec:fft}}

The method that we use here is based on a variational formulation of contact problems given by Kalker~\cite{kalker1977variational} 
and on a simple relationship between a wavy pressure $p(x,y)$ profile and the resulting wavy vertical displacement profile $u^z(x,y)$ due to~\cite{westergaard1939jam,johnson1985ijms,stanley1997}, which can be summarized as:
\be
   p(x,y) = p_0 \cos(k_x x) \cos(k_y y) \quad\Leftrightarrow\quad u^z(x,y) = \frac{2 p_0}{E^*\sqrt{k_x^2+k_y^2}} \cos(k_x x) \cos(k_y y),
  \label{eq:p-u}
\ee
where $x,y$ are in-plane coordinates, $E^*$ is the effective elastic modulus of a pair of contacting elastic materials, given by:
$$
  E^* = \frac{E_1E_2}{(1-\nu_1^2)E_2+(1-\nu_2^2)E_1},
$$
where $E_i,\nu_i$ are Young's moduli and Poisson's ratios of materials $i=1,2$, respectively.
Relationship~\eqref{eq:p-u} is easy to generalize for an arbitrary superposition of modes using a discrete Fourier transformation $\mathcal F$~\cite{stanley1997}:
\be
  [u^z] = \mathcal F^{-1} \left\{\, [w] \,:\, \mathcal F([p]) \,\right\},
  \label{eq:u-p:fft}
\ee
where by colon we denote a double scalar product, and
$$
[u^z] = u^z_{ij}\vec e_i\otimes\vec e_j,\qquad [p] = p_{ij}\vec e_i\otimes\vec e_j
$$ 
are $N\times N$ matrices of vertical displacements and pressures, respectively, defined on a discrete grid $\{x_i,y_j\}$, for $i,j \in [0,N-1]$, 
with $N$ be the number of grid points per period $L$ and $\vec e_i$ is an orthonormal vector basis of dimension $N$, i.e. $\vec e_i\cdot\vec e_j = \delta_{ij}$, where $\delta_{ij}$ is the Kroneker symbol.
By $[w]$ we denote an $N\times N\times N\times N$ matrix of the form:
$$
 [w] =  w_{kl} \delta_{ki}\delta_{lj}\hat{\vec e}_k \otimes\hat{\vec e}_l  \otimes\hat{\vec e}_i \otimes\hat{\vec e}_j,
$$
where $\hat{\vec e}_i$ is another orthonormal basis in Fourier space such that $\mathcal F(p_{ij}\vec e_i\otimes\vec e_j) = \hat p_{ij}\hat{\vec e}_i\otimes\hat{\vec e}_j$  and 
$$
w_{kl} = \frac{L}{\pi E^* \sqrt{k^2+l^2}}
$$

A frictionless non-adhesive normal contact between two linearly elastic solids and roughness $z_1(x,y)$ and $z_2(x,y)$ can be mapped on a contact problem between a rigid rough surface with an effective roughness $z = z_2-z_1$ and an elastic half space with an effective elastic modulus $E^*$ introduced above. 
To use linear elastic approximation and the boundary element method, we assume that the roughness slope remains everywhere vanishingly small $|\nabla z| \ll 1$.
The problem can be formulated as a constrained optimization problem:
\be
  \min\limits_{[p]} \mathbb F, \quad \mathbb F = \frac 12 [p]:[u] + [p]:[g],\mbox{ under constraints } \frac{1}{N^2}\sum p_{ij} = p_0,\;\; p_{ij} \ge 0
  \label{eq:min}
\ee
where $p_0$ is the applied pressure and $[g] = g_{ij}\vec e_i\otimes\vec e_j$ is the matrix of gaps $g_{ij} = z_{ij} - u_{ij} + z_0$, where $z_0$ is an offset value.
This optimization problem can be solved by different optimization techniques~\cite{bertsekas1999nonlinear,bemporad2015optimization}. 
Here we use the methodology suggested in~\cite{polonsky1999numerical}. 
The solution of the optimization problem is a matrix of pressures $[p*]$, which takes positive values in contact zones, where the gap vanishes $g=0$, and zero values $p=0$ where gap is positive. 
At solution, the orthogonality of the gap and pressure is ensured with a given precision $\varepsilon$:
$$
\frac{[p]:[g]}{N^2 p_0 \sqrt{ \langle z^2\rangle}} \le \varepsilon,
$$
where $\sqrt{\langle z^2\rangle}$ is the root mean squared roughness.
Thus, in average we approximately satisfy Hertz-Signorini-Moreau contact conditions~\cite{Wriggers2006} at every grid point:
$$
  p \ge 0,\quad g \ge 0,\quad pg = 0.
$$

\subsection{Computation of the contact area \label{ssec:contact_area}}

\begin{figure}[htb!]
 \centering\includegraphics[width=0.5\textwidth]{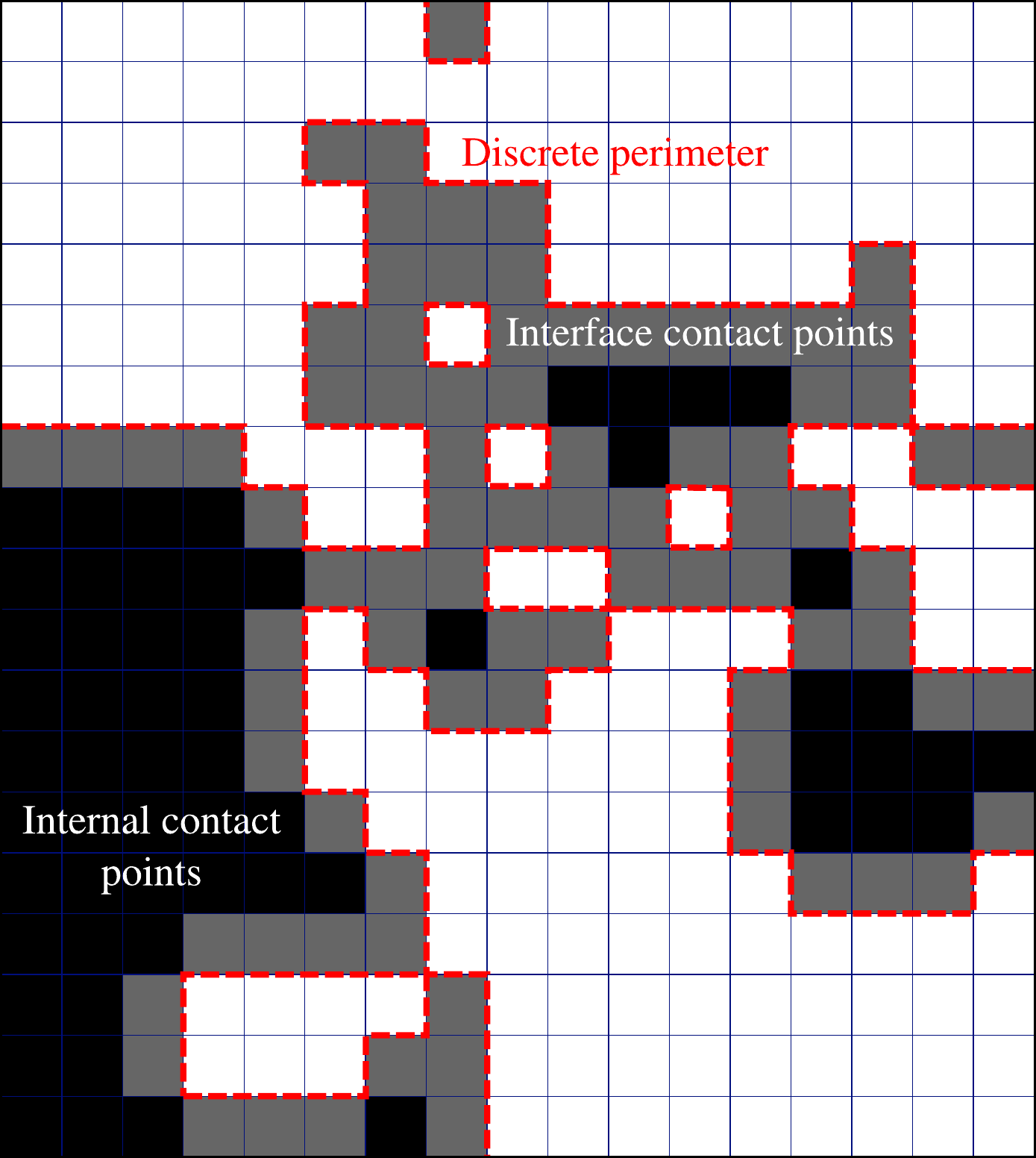}
 \caption{\label{fig:discrete_area}Part of contact area extracted from numerical simulation obtained on a coarse grid. All non-contacting grid points are white cells, contacting grid points are represented by grey and black cells for boundary and interior contact points, respectively. By boundary points we mean all contact points possessing at least one non-contacting nearest neighbour-point. All boundary grid points will be subjected to a morphological modification when the mesh is refined whereas 
 all interior points have a high probability to preserve their contact area (an exception consists in local openings, which can happen under mesh refinement~\cite{dapp2014jpcm,bemporad2015optimization}).
The contact perimeter computed on such a discrete mesh is marked with a dashed line.}
\end{figure}

To compute the discrete contact area $A_d$ we simply count the number of grid points $N_a$ with positive pressure $p_{ij}>0$, each point contributing an area $\Delta x^2$, where $\Delta x=L/N$, $L$ is the side length and $N$ is the number of grid points per side.
The ratio $A_d/A_0 = N_a/N^2$ is then the approximate contact area fraction, since $A_0 = N^2\Delta x^2$ is the nominal contact area. The grid points in contact are split into two types: internal points $N_a^i$, which do not have non-contact grid points in their nearest neighbourhood and boundary contact points $N_a^b$, which have non-contact points in their nearest neighbourhood; $N_a = N_a^i + N_a^b$. We suppose that the true (obtained for $N\to\infty$) boundary between contact and non-contact zones is located somewhere in
between of these discrete measurements. Thus the true contact area $A_*$ can be estimated as
$$
 \frac{A_*}{A_0} \approx \frac{N_a^i + (1-\beta) N_a^b}{N^2} = \frac{N_a}{N^2} - \beta \frac{N_a^b}{N^2},
$$
where $0 < \beta < 1$.
The number of boundary contact points $N_a^b$ correlates directly with the perimeter of contact/non-contact boundaries $S_d = M \Delta x$, where $M$ is the number of switches from contact to non-contact and vice versa along grid points both in vertical and in horizontal directions (Fig.~\ref{fig:discrete_area}). Since for a finite $k_s$ and in the limit of zero grid spacing $N\to\infty$ the contact clusters are smooth figures, the discrete perimeter $S_d$ should be converted into a smooth perimeter $S$, which for an arbitrary isotropic shape can be obtained as
$$
  S =  \frac\pi4 S_d.
$$
The factor $\pi/4$ appears in the conversion from a discrete perimeter measurement to a continuous one because a straight line inclined by an arbitrary angle to a regular Cartesian grid has in average a true length which is $4/\pi$ times shorter than the length computed on a Cartesian grid with vertical and horizontal lines\footnote{In~\cite{pastewka2014sticky} for the same correction the authors used a slightly different factor $4\mathrm{arcsinh}(1)/\pi \approx 1.1222$, whereas our factor $4/\pi \approx 1.2732$.}. It is similar to the conversion of $L_1$ (or Manhattan) metrics to $L_2$ (or Euclidean) metrics. To give an example, a circle with radius $R$ has a discrete perimeter $S_d \approx 8R$, which is $4/\pi$ times bigger than the true perimeter $S = 2\pi R$. Finally, the true contact area fraction can be estimated as
\be
 \frac{A_*}{A_0} \approx \frac{A_d}{A_0} - \beta \frac{\pi}{4} \frac{S_d \Delta x}{A_0},
 \label{eq:true_area}
\ee
where the factor $\beta$ will be determined in the following section. An alternative formulation of this correction can be found in~\cite[Sect.9]{yastrebov2015ijss}.
Since $S_d$ is almost independent of $\Delta x$, and thus $S_d \Delta x \to 0$ as $\Delta x\to0$, from Eq.~\eqref{eq:true_area} it follows that
$$
  \frac{A_d}{A_0} \xrightarrow[\Delta x\to0]{} \frac{A_*}{A_0}.
$$
Note that the value of the correction $\beta \pi S_d \Delta x/4A_0$, which is nothing but the absolute error, is proportional to discretization step $\Delta x$.

\subsection{Corrective factor \label{ssec:factor}}

\begin{figure}[htb!]
 \centering\includegraphics[width=1\textwidth]{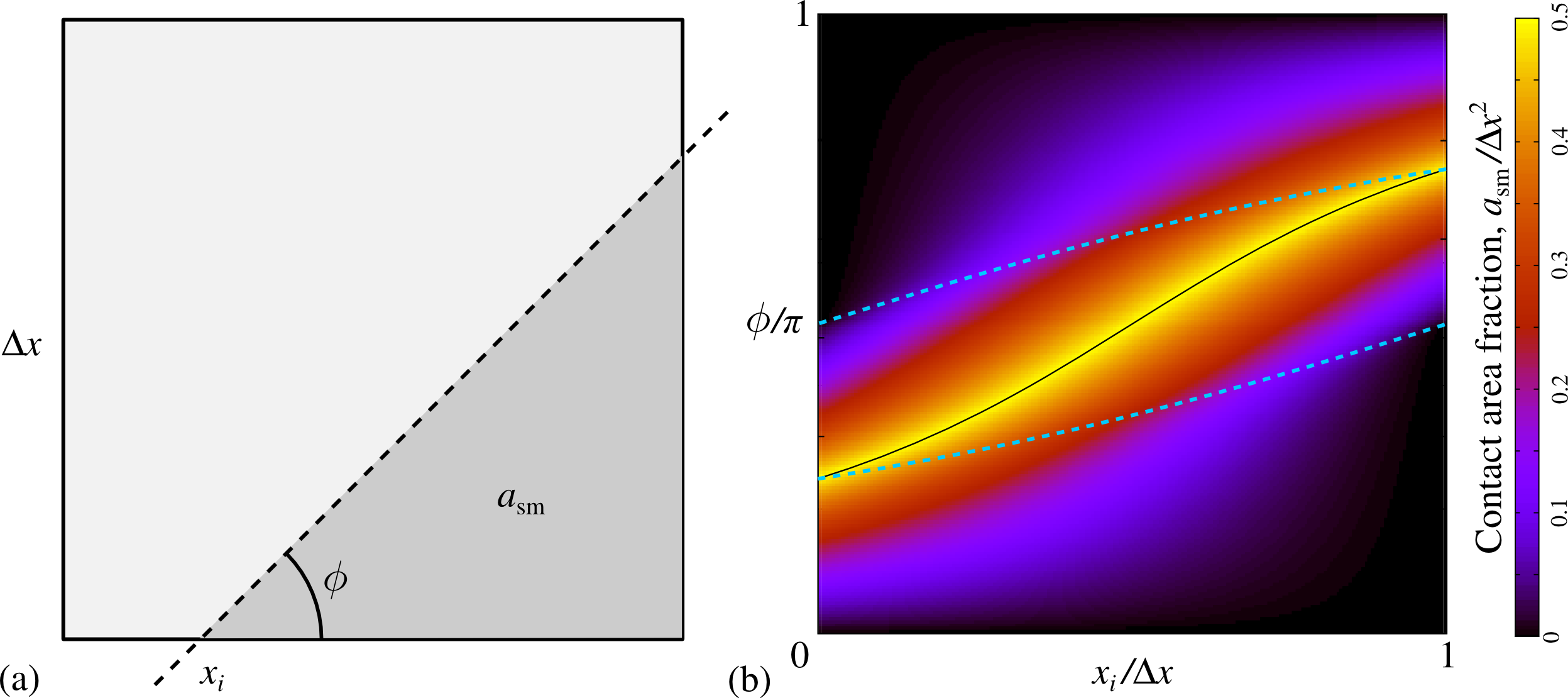}
 \caption{\label{fig:1}Area of the smallest part of a square cut by a straight line: (a) schematic, (b) graphical representation of the integrand of~\eqref{eq:asm_p}, on the abscissa we plot the normalized coordinate of the line intersection with a bottom side of the square $x_i/\Delta x$ and on the ordinate we plot the normalized inclination angle $\phi/\pi$, the colour represents the normalized area of the smallest cut part $a_{\mbox{\tiny sm}}/\Delta x^2$. Solid black line in the centre corresponds to $a_{\mbox{\tiny sm}}=\Delta x^2/2$ and follows $\phi = \atan(\Delta x/(\Delta x-2x_i)$ curve, dashed line depicts $\phi = \atan(\Delta x/(\Delta x-x_i))$ curve.}
\end{figure}

Let us take a square with side $\Delta x$ and cut it into two parts (left and right) by a randomly oriented straight line.
Let us make the following assumptions without loss of generality (see Fig.~\ref{fig:1}(a)):
\begin{enumerate}
  \item Let us consider a square occupying $x \in [0,\Delta x]$, $y \in [0,\Delta x]$.
  \item Let one of the intersections $\{x_i,y_i\}$ of the cut line with the sides of the square lie on the side with $y=0$, so that $x_i \in [0,\Delta x]$, $y_i = 0$ with a uniform probability distribution $P_x(x_i) = 1/\Delta x$.
  \item The angle of inclination\footnote{A seemingly simpler approach would be to introduce a cut position $x$ on one side of the square and a cut position $y$ on another side, but in this case the probability density of the second cut position depends on the first cut and thus cannot be considered uniform which would complicate the following analysis.} of the line with respect to axis $0X$ is denoted $\phi \in [0,\pi]$ with a uniform probability distribution $P_\phi(\phi) = 1/\pi$.
\end{enumerate}

The area of the smaller part $a_{\mbox{\tiny sm}}$ can be easily computed for any given $x_i$ and $\phi$, the remaining (bigger) part has area $\Delta x^2 - a_{\mbox{\tiny sm}}$:
\bes\displaystyle
 a_{\mbox{\tiny sm}}(x_i,\phi) = 
     \begin{cases}
       \frac12(\Delta x-x_i)^2 \tan(\phi),      		&\mbox{ if }  0                \le \phi < \atan(\Delta x/(\Delta x-x_i))\\
        \Delta x(\Delta x-x_i) - \frac{\Delta x^2}{2\tan(\phi)},    	&\mbox{ if }  \atan(\Delta x/(\Delta x-x_i)) \le \phi < \atan(\Delta x/(\Delta x-2x_i)) \mbox{ and } x < \Delta x/2\\
        \Delta xx_i + \frac{\Delta x^2}{2\tan(\phi)},        		&\mbox{ if }  \atan(\Delta x/(\Delta x-2x_i)) \le \phi < \pi/2 \mbox{ and } x < \Delta x/2\\
        \Delta x(\Delta x-x_i) - \frac{\Delta x^2}{2\tan(\phi)},    	&\mbox{ if }  \atan(\Delta x/(\Delta x-x_i)) \le \phi < \pi/2 \mbox{ and } x \ge \Delta x/2\\
        -x_i^2 \tan(\phi),                 			&\mbox{ if }  \phi > \pi - \atan(\Delta x/x_i)\\
        \Delta xx_i + \frac{\Delta x^2}{2\tan(\phi)},        		&\mbox{ if }  \pi - \atan(\Delta x/(2x_i-\Delta x)) \le \phi < \pi - \atan(\Delta x/x_i) \mbox{ and } x \ge \Delta x/2\\
        \Delta x(\Delta x-x_i) - \frac{\Delta x^2}{2\tan(\phi)},    	&\mbox{ if }  \pi/2 \le \phi < \pi - \atan(\Delta x/(2x_i-\Delta x)) \mbox{ and } x \ge \Delta x/2\\
        \Delta xx_i +  \frac{\Delta x^2}{2\tan(\phi)},        		&\mbox{ if }  \pi/2 \le \phi < \pi - \atan(\Delta x/x_i)  \mbox{ and } x < \Delta x/2\\
        \end{cases}
\ees

The mean area of the smaller part can be computed as follows
\be
\langle a_{\mbox{\tiny sm}} \rangle = 
\int\limits_{0}^{\Delta x}\int\limits_{0}^{\pi} a_{\mbox{\tiny sm}}(x_i,\phi) P_x(x_i)P_\phi(\phi) \,d\phi\,dx_i=
\frac{1}{\pi \Delta x}\int\limits_{0}^{\Delta x}\int\limits_{0}^{\pi} a_{\mbox{\tiny sm}}(x_i,\phi) \,d\phi\,dx_i
 \label{eq:asm_p}
\ee
The integrand of~\eqref{eq:asm_p} is plotted in Fig.~\ref{fig:1}(b) and can be evaluated analytically in closed form (see Appendix~\ref{app:factor}) as:
\be
\frac{\langle a_{\mbox{\tiny sm}} \rangle}{\Delta x^2} = \frac{\pi-1+\ln2}{6\pi} \approx 0.150387619
 \label{eq:asm_solve}
\ee
We assume that in our numerical code we overestimate the contact area at border grids by this smaller portion of a square, so
replacing factor $\beta$ in~\eqref{eq:true_area} by expression in~\eqref{eq:asm_solve} we obtain the ultimate equation for the true contact area estimation:
\be
  \boxed{\;A_* \approx A_{d} -  \frac{\pi-1+\ln2}{24} S_d \Delta x\;}
  \label{eq:final_corrected_area}
\ee

\section{Mesh convergence tests~\label{sec:mesh_conv}}

\subsection{Contact area}

\begin{figure}[htb!]
 \includegraphics[width=1\textwidth]{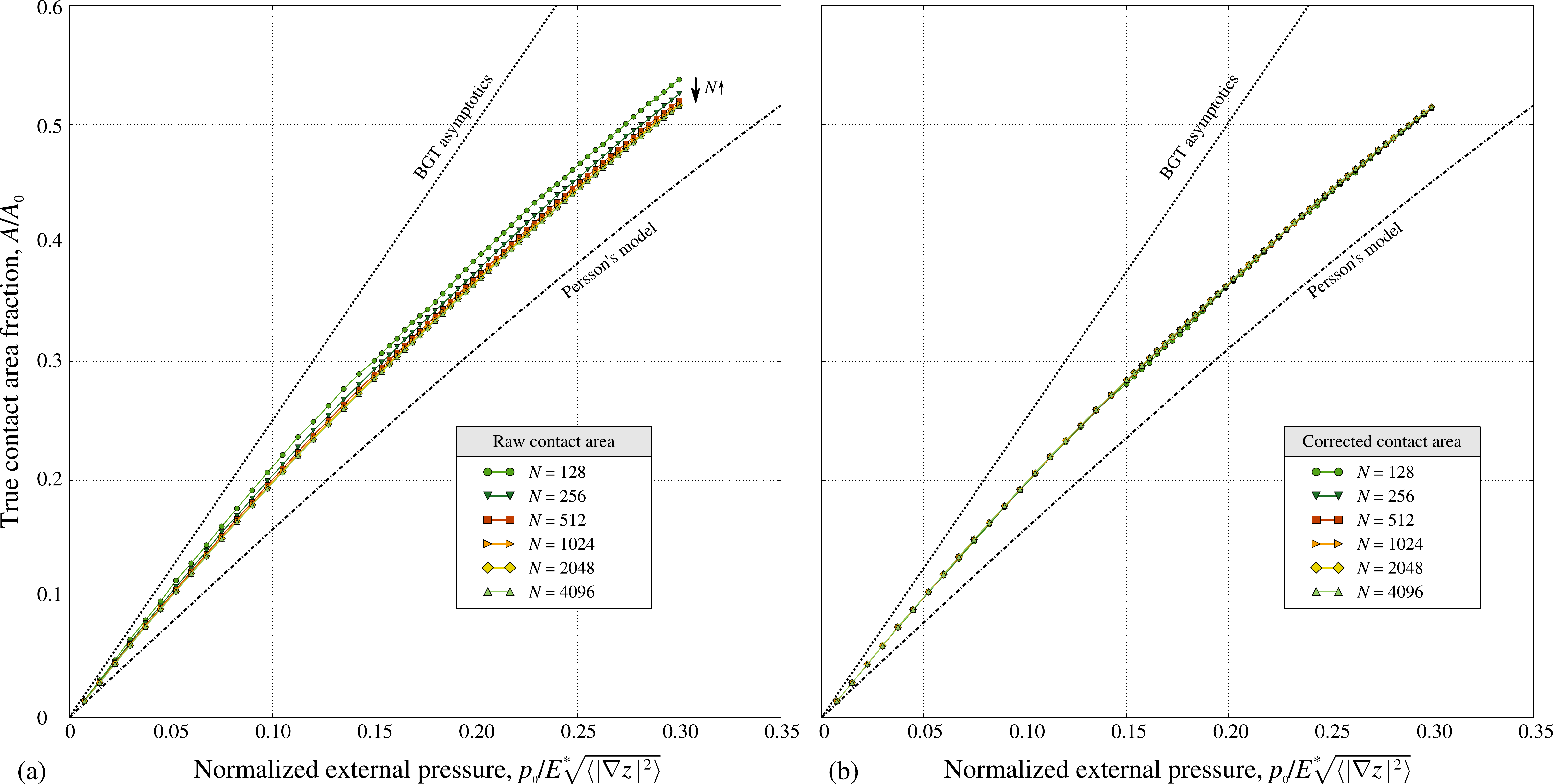}
 \caption{\label{fig:area_ev_1_32}True contact area evolution with applied pressure for a rough surface evaluated for $\tilde k_l = 1$, $\tilde k_s = 32$ (magnification parameter, in Persson's terminology, $\zeta = \tilde k_s/\tilde k_l = 32$), $H=0.8$ with $N= 128, 256, 512, 1024, 2048, 4096$ grid points per side: (a) raw data, (b) corrected data using Eq.~\eqref{eq:final_corrected_area}. 
 Asymptotic solution of asperity based models~\eqref{eq:area_bgt} and Persson's solution~\eqref{eq:area_persson} are plotted for reference purpose with dotted and dash-dotted lines, respectively.}
\end{figure}

\begin{figure}[htb!]
 \includegraphics[width=1\textwidth]{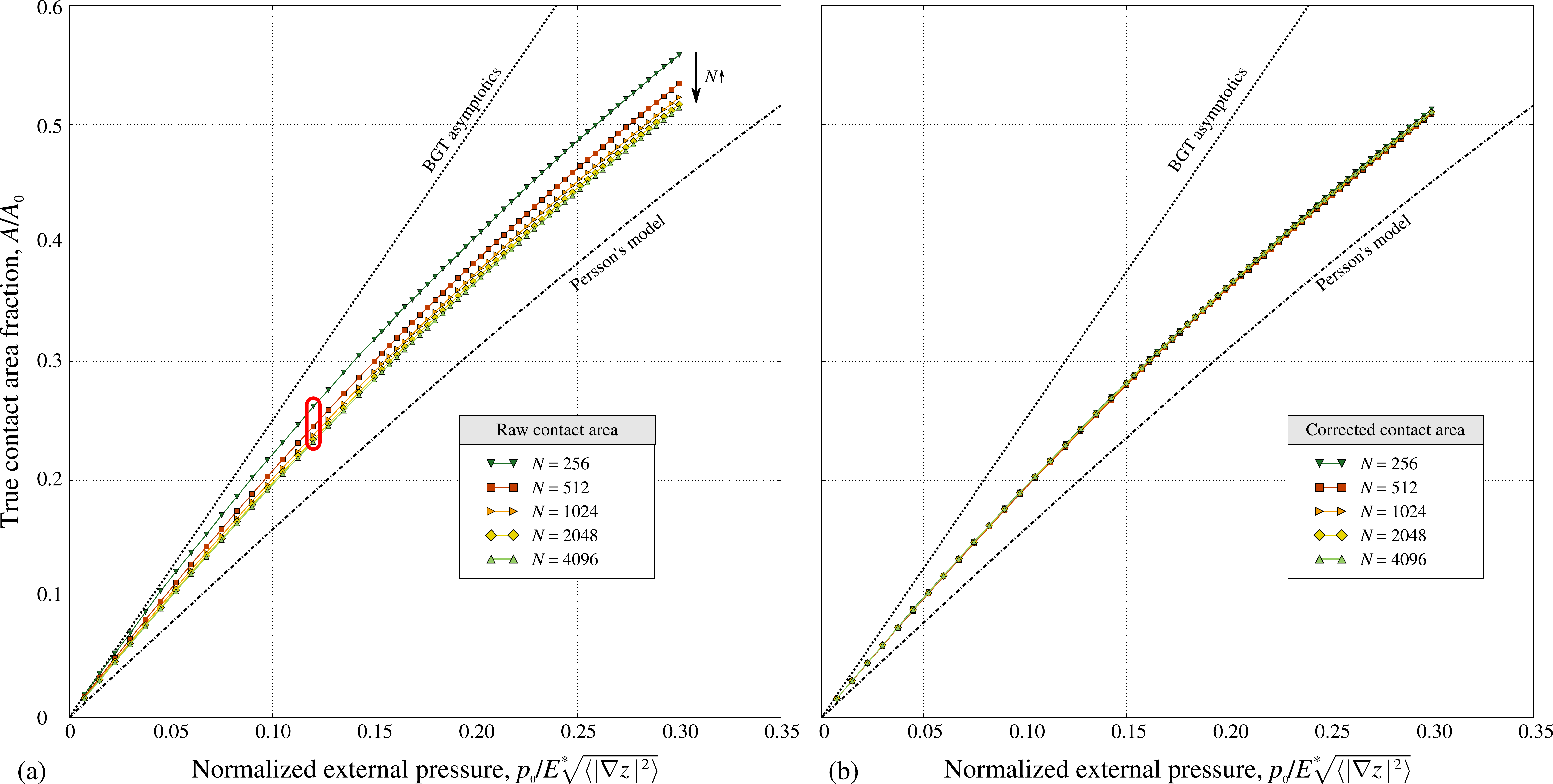}
 \caption{\label{fig:area_ev_4_128}True contact area evolution with applied pressure for a rough surface with a plateau evaluated for $\tilde k_l = 4$, $\tilde k_s = 128$ (magnification $\zeta = 32$), $H=0.8$ with $N= 256, 512, 1024, 2048, 4096$ grid points per side: (a) raw data, (b) corrected data using Eq.~\eqref{eq:final_corrected_area}. Contact area morphologies for points highlighted by a rounded rectangle in (a) are depicted in Fig.~\ref{fig:map_4_128}. Asymptotic solution of asperity based models~\eqref{eq:area_bgt} and Persson's solution~\eqref{eq:area_persson} are plotted for reference purpose with dotted and dash-dotted lines, respectively.}
\end{figure}

To validate our area correction technique, we present a mesh convergence test. We carried out several simulations on surfaces for two Hurst exponents $H=0.4,\;0.8$ and the following combinations of cut-off wavenumbers\footnote{Here, by wavenumber we mean a normalized dimensionless wavenumber $\tilde k = L/\lambda$.} $\{\tilde k_l,\tilde k_s\} = \{1, 32\}, \{1, 64\}$ with discretizations $N\times N$ with $N = 128, 256, 512, 1024, 2048, 4096$, and also for  $\{\tilde k_l,\tilde k_s\} = \{4, 128\}, \{4, 256\}$ with discretizations $N = 256, 512, 1024, 2048, 4096$. Coarse roughness discretizations were obtained from a fine discretization $N=8192$ by sampling every $2, 4, 8, 16$ or $32$ points. 
In all considered cases, the formula~\eqref{eq:final_corrected_area} for corrected true contact area gives accurate results independent of discretization. This suggests that simulations of non-adhesive contact can be done on a rough surface with a very coarse discretization $N = 2\tilde k_s = 2L/\lambda_s$ but with almost the same accuracy(!) as for $N = 32L/\lambda_s$. 
This represents a gain factor of $16^2=256$ in terms of discretization points. It is worth mentioning that the accuracy can be ensured solely for the estimation of the global true contact area but not for other important quantities such as local pressure or gap distribution, which would still require fine discretizations or different corrective techniques.
This correction will allow to perform \emph{without loss of accuracy} rough contact simulations with very broad surface spectra (high Nayak's parameter~\cite{nayak1971tasme}) encountered in real engineering surfaces. To demonstrate the performance of our technique, we show the contact area evolution in Fig.~\ref{fig:area_ev_1_32} 
((a) raw data, (b) corrected data using Eq.~\eqref{eq:final_corrected_area}) for $H=0.8$, $\tilde k_l = 1$, $\tilde k_s = 32$ and different discretizations;
results for $H=0.8$, $\tilde k_l = 4$, $\tilde k_s = 128$ are depicted in Fig.~\ref{fig:area_ev_4_128}.

According to analytical models based on asperity considerations~\cite{greenwood1966prcl,bush1975w,mccool1986w,thomas1999b,greenwood2006w,carbone2008jmps} or based on the evolution of the probability density of contact pressures under increasing magnification, an idea developed by Persson~\cite{persson2001jcp,persson2006contact,manners2006w,dapp2014jpcm}, the contact area fraction $A/A_0$ is proportional to the applied pressure $p_0$ at infinitesimal pressure $p_0$. More precisely, the relation stands as:
\be
  \frac A A_0 = \kappa \frac{p_0}{E^*\sqrt{\langle|\nabla z|^2\rangle}},
  \label{eq:area_bgt}
\ee
where $\kappa$ is the coefficient of proportionality and $\sqrt{\langle|\nabla z|^2\rangle}$ is the root mean squared roughness gradient.
The coefficient $\kappa$ was found to be $\sqrt{2\pi}$ in all asperity based models (which use Nayak's random process model for roughness description~\cite{bush1975w,mccool1986w,thomas1999b,greenwood2006w,carbone2008jmps}), and was deduced to be $\sqrt{8/\pi}$ in Persson's model. 
In Figs.~\ref{fig:area_ev_1_32} and \ref{fig:area_ev_4_128}, to provide bounds, we plot also the asymptotic solution~\eqref{eq:area_bgt} of asperity based models and Persson's solution for the contact area evolution given by:
\be
  \frac{A}{A_0} = \mathrm{erf}\left(\frac{\sqrt2 p_0}{E^*\sqrt{\langle|\nabla z|^2\rangle}}\right).
  \label{eq:area_persson}
\ee
Note that in Fig.~\ref{fig:area_ev_4_128} the results are obtained for a surface with a plateau in the PSD similar to the one shown in Fig.~\ref{fig:rough}(c).

It is important to remark that to carry out this convergence study, the measure of the rms surface gradient should be done using the following equivalence:
$$
\sqrt{\langle|\nabla z|^2\rangle} = \sqrt{2m_2}, 
$$
where $m_2$ is the second spectral moment~\cite{nayak1971tasme,greenwood2006w}.
Since the spectrum and thus the power spectral density of every surface with a cut-off wavenumber $\tilde k_s < N/2$ is independent of the discretization $N$, by measuring the rms surface gradient as the average second spectral moment $2m_2 = m_{02} + m_{20}$ we can find the exact value of $\sqrt{\langle|\nabla z|^2\rangle}$ in the continuous limit. Therefore we can ensure the proper normalization of the pressure. If the rms surface gradient is measured on the real space geometry using forward finite difference (FFD) or central finite difference (CFD) schemes, it becomes a discretization dependent parameter~\cite{greenwood1984unified,paggi2010w} and thus the convergence study on different grids becomes ambiguous as both the normalized pressure and the contact area change simultaneously.

\subsection{Contact area slope}

To demonstrate the accuracy of the contact area evaluation using the suggested technique, we plot the derivative of the contact area with respect to the applied normalized pressure $p_0/E^*\sqrt{\langle|\nabla z|\rangle}$. The slope denoted $\kappa(p_0)$ is evaluated as a finite difference 
\be
  \kappa_i = \kappa(p^i_0) = E^*\sqrt{\langle|\nabla z|^2\rangle} \left.\frac{\partial A/A_0}{\partial p_0}\right|_{p_0^i} \approx  \frac{E^*\sqrt{\langle|\nabla z|^2\rangle}}{A_0} \frac{A^{i+1/2} - A^{i-1/2}}{p_0^{i+1/2}-p_0^{i-1/2}}.
  \label{eq:kappa}
\ee
Defined in this way, the slope should tend to the area-pressure proportionality coefficient as pressure tends to zero:
$$
\lim\limits_{p_0\to0} \frac{E^*\sqrt{\langle|\nabla z|^2\rangle}}{A_0}\frac{\partial A}{\partial p_0} = \kappa_0.
$$
This proportionality coefficient $\kappa_0$, predicted by asperity based model~\cite{bush1975w,thomas1999b,mccool1986w,greenwood2006w,carbone2008jmps} to be $\sqrt{2\pi}$ and
by Persson's model~\cite{persson2001jcp} to be $\sqrt{8/\pi}$, was a topic of numerous numerical studies~\cite{hyun2004pre,hyun2007ti,paggi2010w,campana2007epl,putignano2012jmps,yastrebov2012pre,pohrt2012prl,prodanov2014tl,yastrebov2015ijss}.
To evaluate the slope correctly we use the corrected contact area and thus compute it as follows (the upper index $c$ indicates the corrected value):
\be
  \kappa^c = \frac{E^*\sqrt{\langle|\nabla z|^2\rangle}}{A_0} \frac{\partial A_c}{\partial p_0} =
  \frac{E^*\sqrt{\langle|\nabla z|^2\rangle}}{A_0} \left(\frac{\partial A}{\partial p_0} -\frac{(\pi - 1 +\ln2)\Delta x}{24}\frac{\partial{S }}{\partial p_0}\right)
  \label{eq:kappa_c}
\ee
To slightly reduce oscillations we smooth the slope using:
\be
  \bar\kappa_i = \frac12\kappa_i + \frac14\left(\kappa_{i-1} + \kappa_{i+1}\right).
  \label{eq:kappa_bar}
\ee

\begin{figure}[htb!]
 \includegraphics[width=1\textwidth]{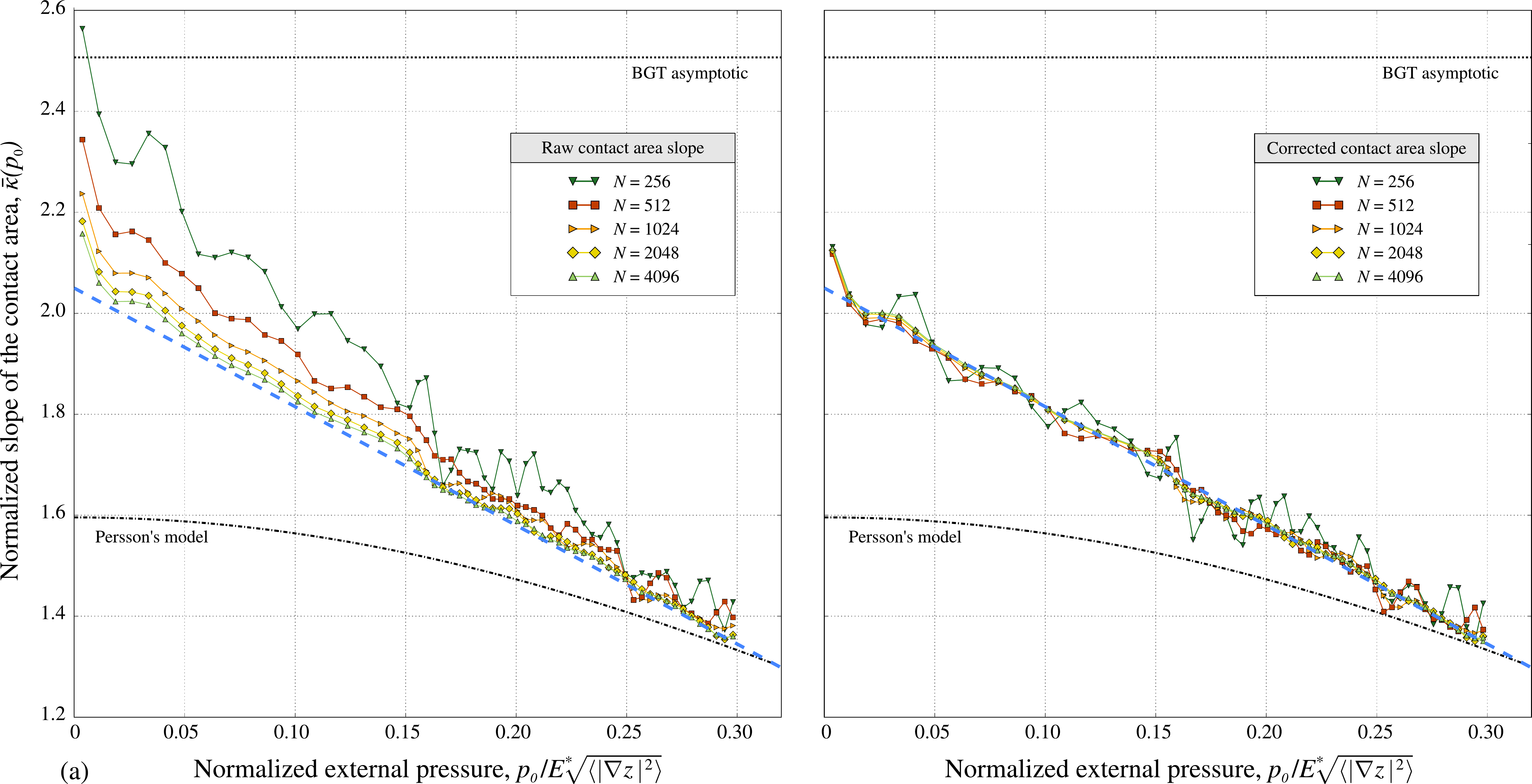}
 \caption{\label{fig:slope_4_128_h08}Slope of the true contact area evolution with applied pressure for a rough surface with a plateau evaluated for $\tilde k_l = 4$, $\tilde k_s = 128$ (magnification $\zeta = 32$), $H=0.8$ with $N= 256, 512, 1024, 2048, 4096$ grid points per side: (a) raw data, (b) corrected data using Eq.~\eqref{eq:final_corrected_area}. Linear curve (dashed blue line), which fits the corrected data, is given by $\kappa = 2.05 - 2.35 p_0/E^*\sqrt{\langle|\nabla z|^2\rangle}$.}
\end{figure}

\begin{figure}[htb!]
 \includegraphics[width=1\textwidth]{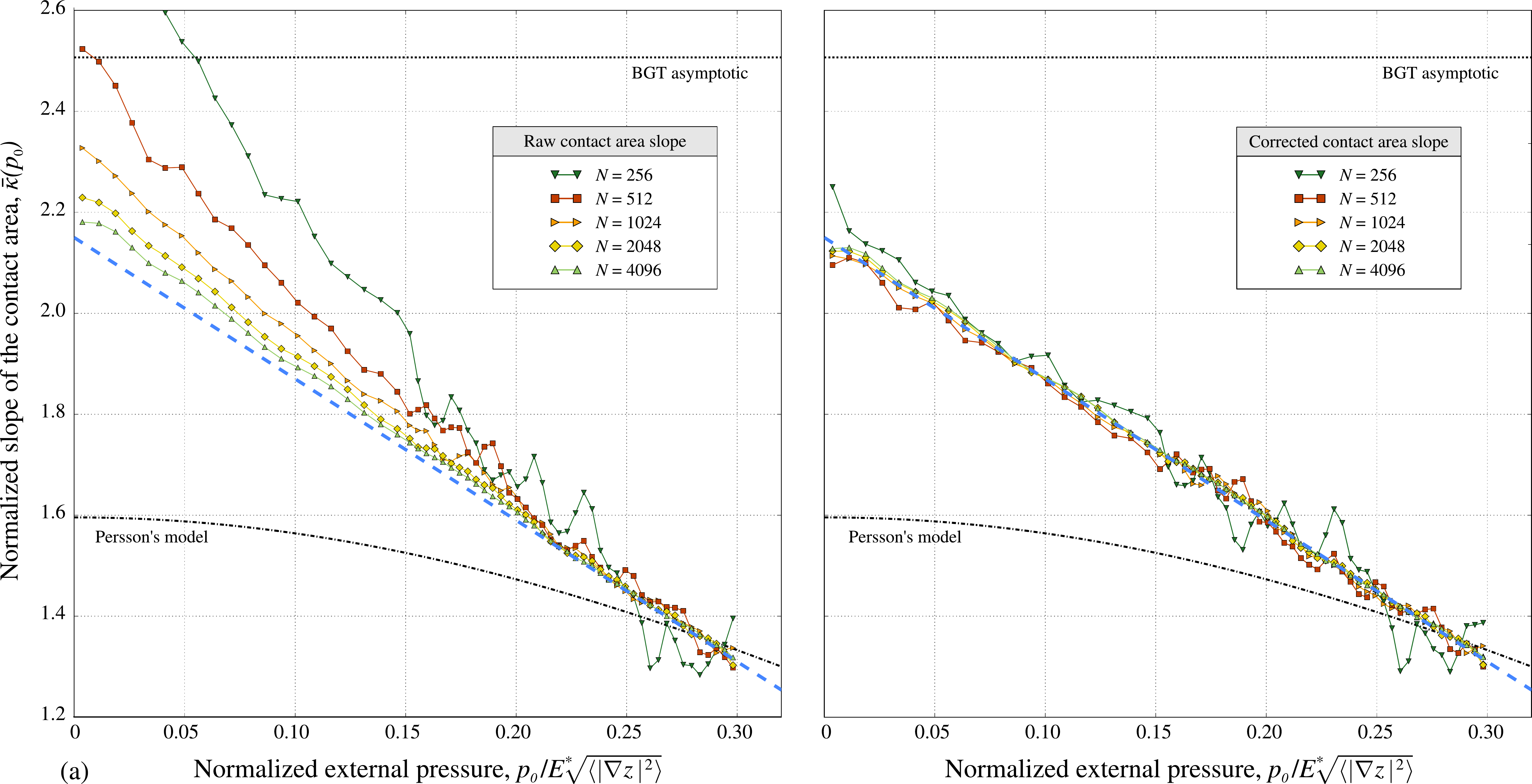}
 \caption{\label{fig:slope_4_128_h04}Slope of the true contact area evolution with applied pressure for a rough surface with a plateau evaluated for $\tilde k_l = 4$, $\tilde k_s = 128$ (magnification $\zeta = 32$), $H=0.4$ with $N= 256, 512, 1024, 2048, 4096$ grid points per side: (a) raw data, (b) corrected data using Eq.~\eqref{eq:final_corrected_area}.
 Linear curve (dashed blue line), which fits the corrected data, is given by $\kappa = 2.15 - 2.80 p_0/E^*\sqrt{\langle|\nabla z|^2\rangle}$.}
\end{figure}

In Figs.~\ref{fig:slope_4_128_h08} and \ref{fig:slope_4_128_h04} the slopes are computed for $\tilde k_l = 4$, $\tilde k_s = 128$ for $H=0.8$ and $H=0.4$, respectively and for different grids with $N= 256, 512, 1024, 2048, 4096$ using raw data with Eq.~\eqref{eq:kappa} and corrected data with Eq.~\eqref{eq:kappa_c}, both results were slightly smoothed using Eq.~\eqref{eq:kappa_bar}. The results are compared with the slope of Persson's model~\cite{persson2001jcp} given by
\be
  \kappa_P(p_0) = \frac{\partial\,\mathrm{erf}(\sqrt2 p')}{\partial p'} = \sqrt{\frac8\pi}\exp\left[-\frac{p_0^2}{E^{*2} \langle|\nabla z|^2\rangle}\right],
  \label{eq:persson}
\ee
where $p' = p_0/E^*\sqrt{\langle|\nabla z|^2\rangle}$.
Note that the effect of the area correction is considerable even for the finest discretization $N=4096$ especially in the interval of small pressures.

The corrected data of the slopes can be accurately fitted by a linear function suggesting 
that the contact area evolves approximately as a polynomial of the second degree:
\be
\frac A A_0 \approx \kappa_0 p' - c p'^2.
\ee
For the presented results, $\kappa_0 \approx 2.05$, $c \approx 2.35$ for $\tilde k_l = 4$, $\tilde k_s = 128$, $H=0.8$ and
$\kappa_0 \approx 2.15$, $c \approx 2.80$ for $\tilde k_l = 4$, $\tilde k_s = 128$, $H=0.4$. These values of $\kappa_0$ are comparable with those found in the literature~\cite{hyun2004pre,hyun2007ti,paggi2010w,campana2007epl,putignano2012jmps,yastrebov2012pre,pohrt2012prl,prodanov2014tl,yastrebov2015ijss}. However, a more rigorous statistical analysis would be needed to investigate the separate roles of the fractal dimension (or Hurst exponent) and Nayak's parameter.

\subsection{Morphology of contact clusters\label{sec:morphology}}

\begin{figure}[htb!]
 \includegraphics[width=1\textwidth]{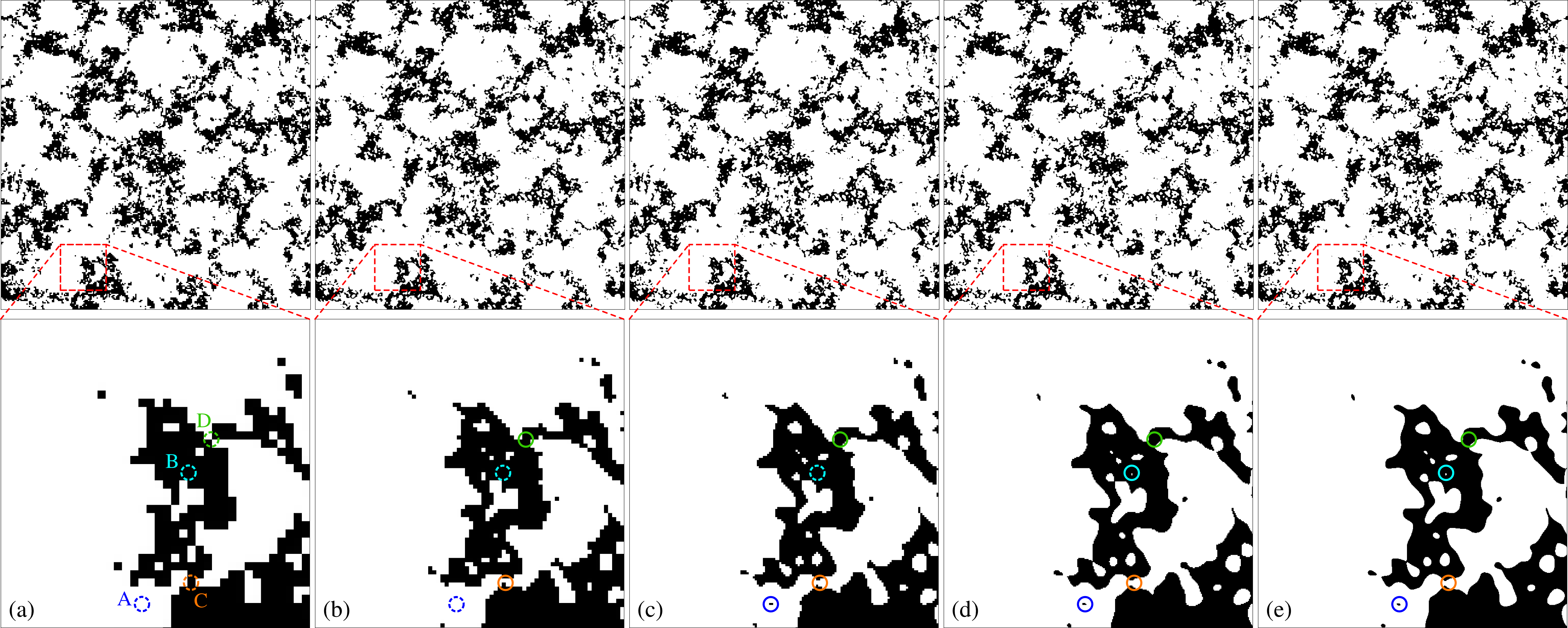}
 \caption{\label{fig:map_4_128}Distribution of contact clusters at pressure $p_0/E^*\sqrt{\langle|\nabla z|\rangle} \approx 0.12$ for a rough surface with a plateau for $\tilde k_l = 4$, $\tilde k_s = 128$ (magnification $\zeta = 32$), $H=0.8$  (it corresponds to the corrected contact area $A'\approx23\;\%$) evaluated with $N = 256, 512, 1024, 2048, 4096$ grid points per side (a)-(e), respectively. A zoom on a small portion demonstrates a mesh convergence to well resolved smooth contact clusters. Some morphological features, which undergo topological changes with mesh refinement, are highlighted (with dashed line, when the true topology is not yet captured and with solid line when it is.)}
\end{figure}

The integral measure of the contact area is an important but not the only characteristic that determines physical properties of contact interfaces. The morphology of contact clusters and their distribution also play an important role in many interfacial phenomena~\cite{greenwood1966ecr,dapp2016fluid}.
In Fig.~\ref{fig:map_4_128} we show maps of contact clusters computed at different discretizations for applied pressure $p_0/E^*\sqrt{\langle|\nabla z|\rangle} \approx 0.12$ giving $A/A_0 \approx 22.5$ \% for $H=0.8$, $\tilde k_l = 4$, $\tilde k_s = 128$. 
Three discretization-dependent morphological effects can be distinguished: 1) loss of small contact spots at coarser grids (marked with A in Fig.~\ref{fig:map_4_128}) (see discussion in~\cite{dapp2014jpcm,bemporad2015optimization}), 2) loss of small non-contact spots at coarser grids (marked with B in Fig.~\ref{fig:map_4_128}), 3) ambiguity in connecting or non-connecting between contact clusters connected by a single grid point (marked with C in Fig.~\ref{fig:map_4_128}) or a single next-nearest connection (marked with D in Fig.~\ref{fig:map_4_128}).
The error 3) is probably the most critical for mass transport analysis in near-percolation regime. However, at reasonable discretizations $N = 4L/\lambda_s$ ($\tilde k_s = N/4$) in most cases this problem  does not persist. Overall, even the coarsest discretization $N = 2L/\lambda_s$ ($\tilde k_s = N/2$) shows a reasonable estimation of contact clusters with most morphological features present like on much finer discretizations.

\begin{figure}[htb!]
 \includegraphics[width=1\textwidth]{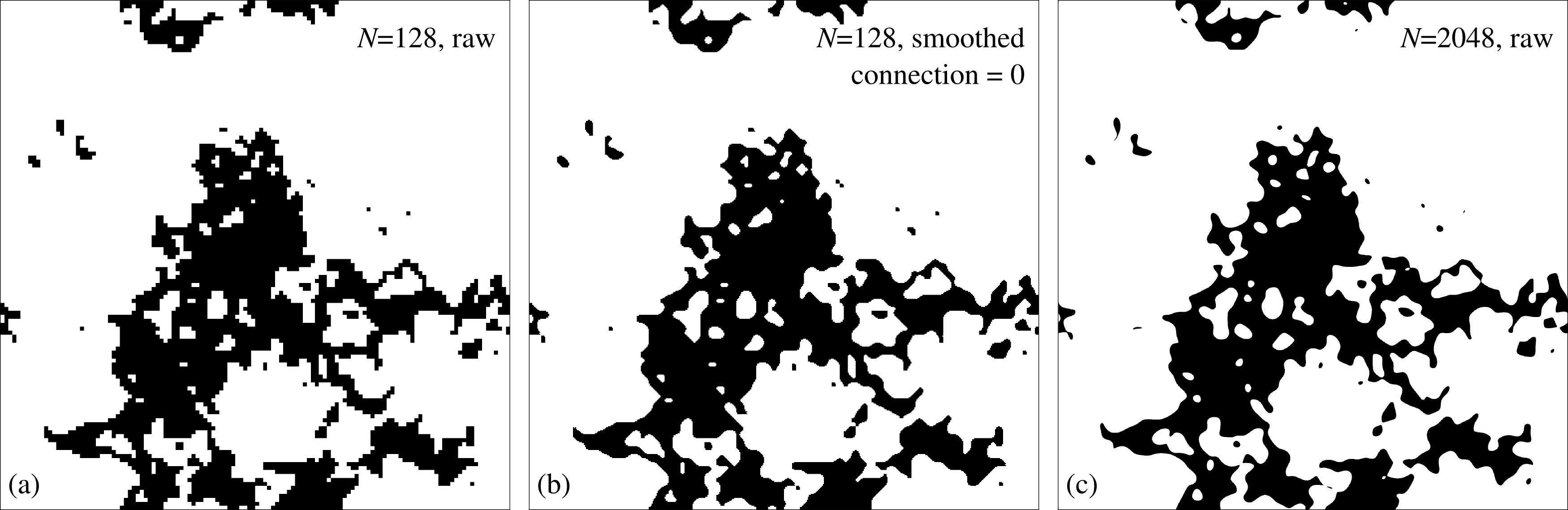}
 \caption{\label{fig:morpho_1_32}Smoothing of contact clusters morphology for $\tilde k_l = 1$, $\tilde k_s = 32$ (magnification $\zeta = 32$), $H=0.8$ at pressure $p_0/E^*\sqrt{\langle|\nabla z|\rangle} \approx 0.12$ and the corrected contact area $A'\approx22\;\%$: (a) coarse grid simulation results for $N=128$, (b) smoothed version of $N=128$ simulations obtained with topological smoothing algorithm~\cite{couprie2004topology}, (c) fine grid simulation result for $N=2048$.}
\end{figure}

\begin{figure}[htb!]
 \includegraphics[width=1\textwidth]{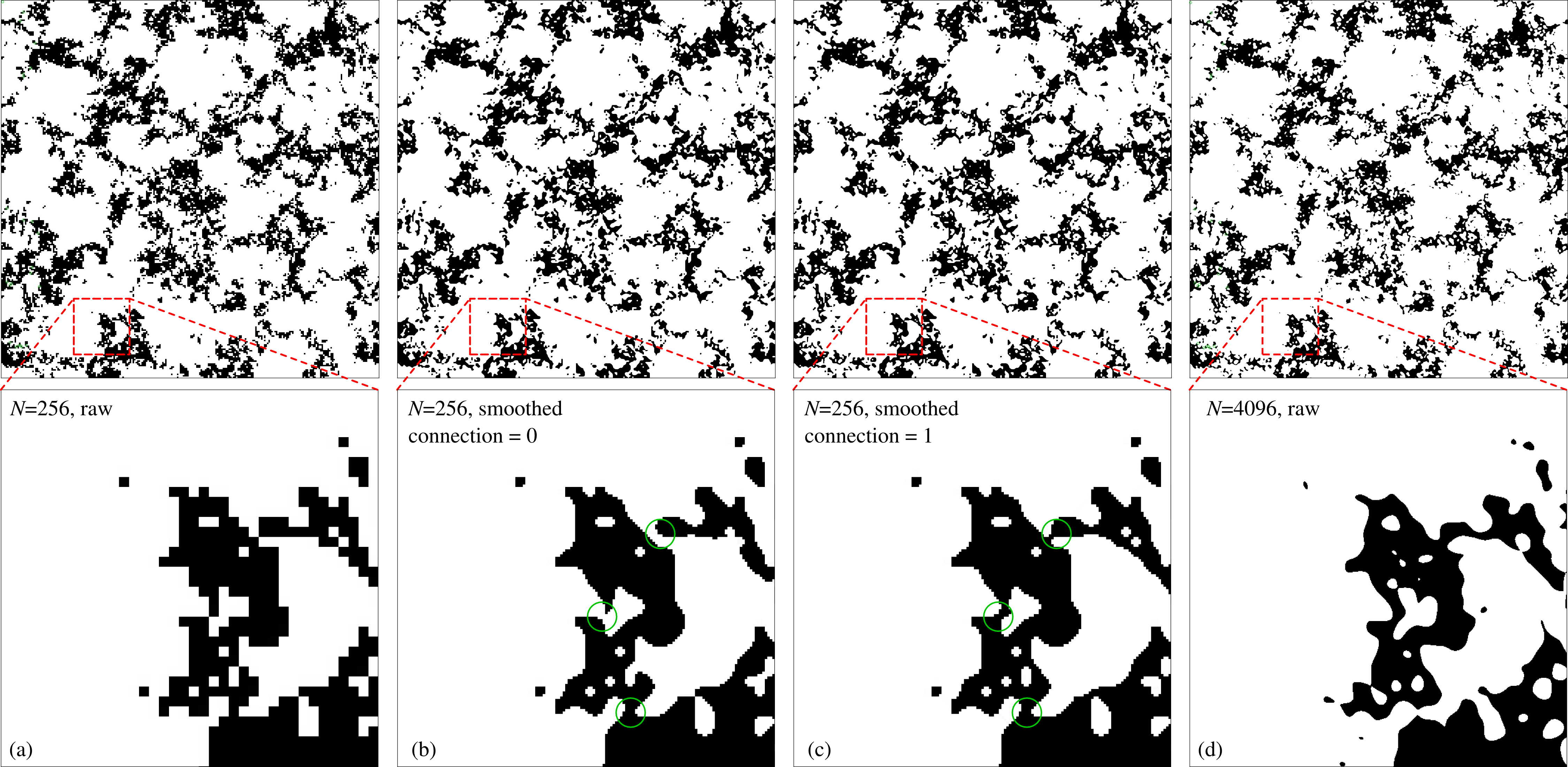}
 \caption{\label{fig:morpho_4_128}Smoothing of contact clusters morphology for $\tilde k_l = 4$, $\tilde k_s = 128$ (magnification $\zeta = 32$), $H=0.8$ at pressure $p_0/E^*\sqrt{\langle|\nabla z|\rangle} \approx 0.12$ and the corrected contact area $A'\approx23\;\%$: (a) coarse grid simulation results for $N=256$, smoothed version of $N=256$ simulations obtained with topological smoothing algorithm~\cite{couprie2004topology} for (b) connnection=0 and (c) connection=1, (d) fine grid simulation result for $N=4096$.}
\end{figure}

To adjust better the morphology of contact clusters obtained with a coarse grid, we use a smoothing algorithm that preserves topology~\cite{couprie2004topology} (see Figs.~\ref{fig:morpho_1_32}, \ref{fig:morpho_4_128}) of the contact cluster map, which is an important feature for rough contact problems. The general method is based on the simultaneous smoothing by metric disks of increasing radius and homotopic cutting/filling, which allow to preserve the topology. For the case of zoomed pixelized images (macropixels of minimal size contain several pixels, i.e. $\times2$ zoom results in $2\times2$ pixels) the method preserves centres of macropixels and smooths the image; the chessboard metric $d(x,y) = \max\{|x_1-y_1|,|x_2-y_2|\}$ is used to construct the topological skeleton of the original image and the Euclidean metric is used for the homotopic filter. All details of the method can be found in~\cite{couprie2004topology}. Moreover, a source code to deal with raster images is provided by the authors of the algorithm on their web-page~\cite{couprie_url}. 

\begin{figure}[htb!]
 \includegraphics[width=1\textwidth]{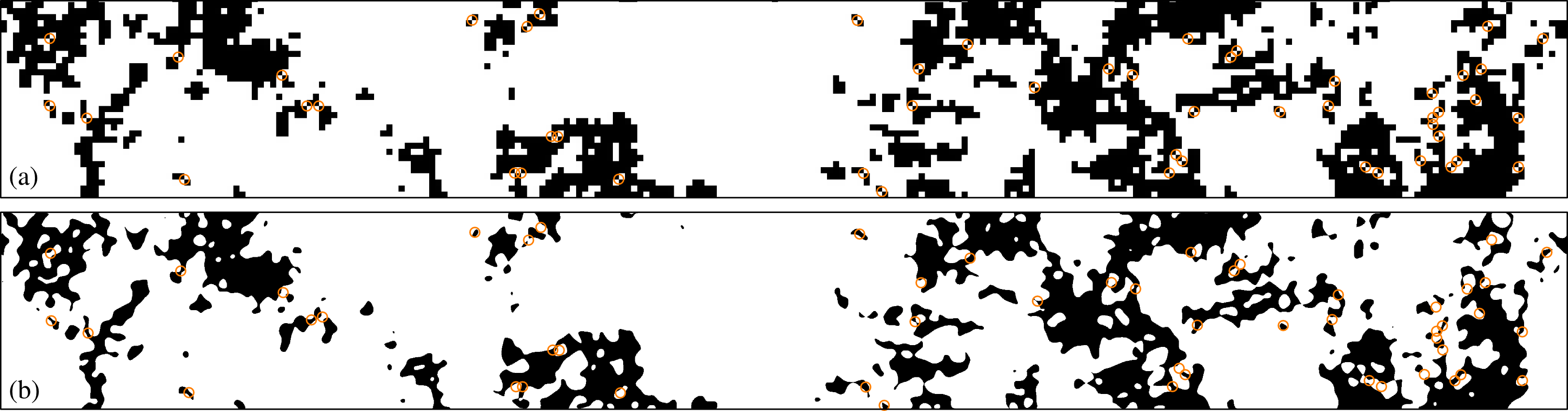}
 \caption{\label{fig:connection}Comparison of a part of results obtained with (a) a coarse grid $N=256$ and (b) a fine grid simulations $N=4096$;  we highlight with circles next-nearest contact pixels connections  in the coarse grid and the corresponding true morphology obtained with the fine grid; the map was computed for $\tilde k_l = 4$, $\tilde k_s = 128$, $H=0.8$.}
\end{figure}

Figs.~\ref{fig:morpho_1_32} and \ref{fig:morpho_4_128} present the results of the smoothing procedure, in which every contact point (black pixel) in the raw data was replaced by $4\times4$ pixels before running the smoothing procedure. The figures correspond to $H=0.8$, $\tilde k_l = 1$, $\tilde k_s = 32$ and $H=0.8$, $\tilde k_l = 4$, $\tilde k_s = 128$, respectively.
Even though such a topology-preserving smoothing cannot resolve the aforementioned problems 1)-3) in cluster computations, it nevertheless delivers improved contact clusters very similar to those obtained with a much finer discretization.
Contact clusters connected only through a next-nearest neighbour connection, can be considered by the algorithm to be topologically disjoint (connection=0) or joint (connection=1) (compare (b) and (c) in Fig.~\ref{fig:morpho_4_128}), which will result in different topological connections critical for the study of percolation. According to our analysis (see example in Fig.~\ref{fig:connection}), it is roughly equivalently probable to find in finely discretized simulations next-nearest neighbours of the coarse mesh to be connected or disconnected. Thus, in percolation analysis assuming connection=0 and =1 could provide us with upper $A_{\mbox{\tiny sm}}^u$ and lower $A_{\mbox{\tiny sm}}^l$ bounds of the percolation limit.
However, an accurate comparison between these bounds and the true contact area revealed that this topological smoothing gives a contact area estimation close to the reference coarse simulation without correction, i.e. strongly overestimates the true contact area given by~\eqref{eq:final_corrected_area}. It follows that morphological smoothing, even if it provides more realistic micro-contact morphologies, cannot be used for quantitative estimation of the true contact area.

\section{Conclusion\label{sec:conclusion}}

In this paper we suggested a correction for the contact area computation in numerical analysis of non-adhesive frictionless contact between elastic rough surfaces. 
Simple geometrical considerations and the evaluation of the perimeter of contact clusters inspired from our previous work~\cite{yastrebov2015ijss} enables us to obtain an accurate estimation of the contact area. The technique is validated using a properly designed mesh convergence test on a particular spectral boundary element method~\cite{stanley1997}, which uses a regular grid combined with conjugate gradient method from~\cite{polonsky1999numerical}. We believe that this approach will be helpful for all spectral based techniques.
We demonstrate on several examples that this correction, which vanishes as the grid spacing tends to zero, ensures an accurate contact area estimation even for a coarse grid with grid spacing $\Delta x = \lambda_s/2$, where $\lambda_s$ is the shortest wavelength present in the roughness spectrum.  
Moreover, the correction enables us to compute accurately the contact area slope. 
These results suggest that when accurately computed, the derivative of the contact area with respect to the contact pressure can be accurately approximated by a linear decreasing function at least up to $50$ \% of contact area, thus the contact area evolution can be properly described by a second order polynomial without a zero-degree coefficient.

In addition to the integral measure of the contact area, we used a topology preserving smoothing technique developed in~\cite{couprie2004topology}, which improves the appearance of contact clusters morphology. 
However, this smoothing does not permit to obtain an accurate quantitative measure of the true contact area.

Overall, using very moderate grid sizes the area correction technique allows to access accurate integral measure of the true contact area in the mechanics of rough contact.
Thus it opens new perspectives in numerical studies of contact properties of rough surfaces.

\vspace{1cm}
\noindent{\bf Acknowledgement}

\noindent Enriching discussions with Valentine Rey and Lucas Fr\'erot are greatly appreciated. 

\appendix

\section{\label{app:factor}Integration of the corrective factor}

To evaluate the size of the smallest part of a square cut by a randomly oriented line we split the integral~\eqref{eq:asm} into four parts:
\be
\begin{split}
&\langle a_{\mbox{\tiny sm}} \rangle = 
\frac{2}{\pi h} \left[
\frac12 \int\limits_{0}^{h}\int\limits_{0}^{\atan\left(\frac{h}{h-x}\right)} (1-x)^2\tan(\phi) \,d\phi\,dx+
\int\limits_{0}^{h/2}\int\limits_{\atan\left(\frac{h}{h-x}\right)}^{\atan\left(\frac{h}{h-2x}\right)} \left\{(1 - x) - \frac{1}{2\tan(\phi)}\right\} \,d\phi\,dx+\right.\\
&
\left.
\int\limits_{0}^{h/2}\int\limits_{\atan\left(\frac{h}{h-2x}\right)}^{\pi/2} \left\{x + \frac{1}{2\tan(\phi)}\right\}\,d\phi\,dx+
\int\limits_{h/2}^{h}\int\limits_{\atan\left(\frac{h}{h-x}\right)}^{\pi/2} \left\{(1 - x) - \frac{1}{2\tan(\phi)}\right\} \,d\phi\,dx
\right] = \frac{2}{\pi h}[I_1+I_2+I_3+I_4]
\end{split}
 \label{eq:asm}
\ee
where integrals can be evaluated separately yielding the following expressions:
\be
I_1 = \frac{h^3}{16}\left[8/3+4/3\ln(2) -2\pi/3\right]
\ee
\be
I_2 = \frac{h^3}{16}\left[\pi + 2\ln(5/4) - 2 + 8\atan(1/2) - 5\atan(4/3)\right]
\ee
\be
I_3 = \frac{h^3}{16}\left[\pi - 2\right]
\ee
\be
I_4 = \frac{h^3}{16}\left[\pi - 2\atan(2) - 2\ln(5/4)\right]
\ee
Which finally gives the following expression:
\be
\langle a_{\mbox{\tiny sm}} \rangle = \frac{\pi-1+\ln2 }{6\pi}h^2
\ee
We used the following relationships:
$$8\,\atan\left(\frac12\right) - 5\,\atan\left(\frac43\right) = -2\,\atan\left(\frac12\right) \quad \mbox{ and }\quad \atan\left(\frac12\right)+\atan(2) = \frac\pi2$$


\end{document}